\documentclass[12pt]{article}
\usepackage{graphicx}

\def \gord{$ \raisebox{-.3ex}{$\stackrel{>}{_{\sim}}$}$}

\def\ignore#1{{}}

\let\oldtheequation=\theequation
\def\doteqs#1{\setcounter{equation}{0}            
\def\theequation{{#1}.\oldtheequation}}
\newcounter{sxn}
\def\sx#1{\addtocounter{sxn}{1} \vskip 1.cm  \goodbreak
\noindent{\large\bf\leftline{\thesxn.~~#1}} \nobreak \vskip -.5cm}
\def\sxn#1{\sx{#1} \doteqs{\thesxn}}

\newcounter{axn}

\date{}

\newdimen\mybaselineskip
\newcommand{\beeq}{\begin{equation}}
\newcommand{\eneq}{\end{equation}}
\newcommand{\beqn}{\begin{eqnarray}}
\newcommand{\eeqn}{\end{eqnarray}}

\def\mybig{\displaystyle \strut }

\def\la{\raise.16ex\hbox{$\langle$}\lower.16ex\hbox{}  }
\def\ra{\, \raise.16ex\hbox{$\rangle$}\lower.16ex\hbox{} }
\def\go{\rightarrow}

\def\onehalf{ \hbox{${1\over 2}$} }

\def\Pl{{\rm Pl}}
\def\eff{{\rm eff}}

\def\ep{\epsilon}
\def\psibar{ \psi \kern-.65em\raise.6em\hbox{$-$} \lower.6em\hbox{} }
\def\psibarb{ \psi \kern-.65em\raise.6em\hbox{$-$}  }
\def\Rbar{{\overline R}}

\def\myfrac#1#2{{\mybig #1\over \mybig #2}}

\begin{document}

\thispagestyle{empty}

\baselineskip=12pt



\vspace*{3.cm}

\begin{center}  
{\LARGE \bf False Vacuum Lumps with Fermions}
\end{center}

\baselineskip=14pt

\vspace{3cm}
\begin{center}
{\bf  Ramin G. Daghigh}
\end{center}

\centerline{\small \it Physics Department, 
University of Winnipeg, Winnipeg, Manitoba R3B 2E9, Canada}



\vspace{3cm}
\begin{abstract}
This paper studies false vacuum lumps surrounded by the true vacuum in a 
real scalar field 
potential in flat spacetime.  Fermions reside in the core 
of the lump, which are coupled with the scalar field 
via Yukawa interaction.
Such lumps are stable against spherical collapse 
and deformation from spherical shape based on energetics considerations.   
The fermions inside the lump are treated as a 
uniform Fermi gas.  We consider the Fermi gas in
both ultrarelativistic and nonrelativistic limits. 
The mass and size of these lumps depend on the scale characterizing 
the scalar field potential as well as the mass density of the fermions.
\end{abstract}

\baselineskip=20pt plus 1pt minus 1pt

\newpage

\sxn{Introduction }


\vskip 1cm

\ignore{
When a scalar field potential has two non-degenerate minima,  the absolute
minimum of the potential corresponds to the true vacuum, while the other 
the false vacuum.  Systems, in which false and true vacua coexist, are of 
great interest.
The universe is full of various structures such as
black holes, stars and dark matter. Does there exist
any structure which has a false vacuum in the middle?
If the entire universe is in the false vacuum, it decays into
the true vacuum through bubble creation by quantum tunneling \cite{Coleman1}.
What happens if a false vacuum core is surrounded by the true 
vacuum \cite{Daghigh, Hosotani2}?
If the size of the core is smaller than a critical radius, the core
would quickly decay with its energy dissipating to space infinity.
If the size of the core is larger than the critical radius, the 
core becomes a black hole.  In either case the configuration cannot
be static.  Fate of false-vacuum bubbles has been extensively discussed
in the literature \cite{Blau, NoGo, Galtsov}.
It has been shown that new configurations, cosmic shells, emerge in
a simple real scalar field theory in which spherical shells of the 
true vacuum
are immersed in the false vacuum \cite{Hosotani1}.  Those static 
cosmic shells
can  exist thanks to gravitational interactions.  However, a 
static ball of the
false vacuum immersed in the true vacuum is not possible.
Recently it has been demonstrated that 
a static false vacuum core becomes possible if there are
fermions coupled to the scalar field \cite{Daghigh1}.  These configurations 
are gravitating fermionic lumps.
The amount of fermions necessary is  modest.  
\ignore{It is much less than the critical amount obtained by simple
energetics. } 
Gravitational interactions as well as Yukawa 
interactions play a key role.  
\ignore {Gravitating lumps are  quite different from Q-balls, boson stars 
or Fermi-balls. In Q-balls  the conserved  charge of the scalar field is 
central to achieve the stability \cite{Qball}.  There is no such charge of 
the scalar field
in our model.  In boson stars gravitational interactions as well as the 
conserved charge play a key role \cite{BosonStar1, BosonStar2}.  The model 
for Fermi balls is very similar to ours \cite{Campbell}.  The false vacuum 
is surrounded by the true vacuum.  In Fermi balls fermions are localized 
in the transition region, or the domain wall, i.e.\ they reside in the 
surface region of the lumps.  In our model fermions reside in the bulk 
region inside the lump and the Yukawa interaction 
becomes essential.}Gravitational interactions can produce lump
solutions in non-Abelian gauge theory as well.  Even in the pure 
Einstein-Yang-Mills theory stable monopole solutions appear in 
the asymptotically anti-de Sitter space \cite{Bjoraker, Bartnik, Torii}.
\ignore {We shall show in the present paper that such gravitational 
lumps appear
even in a simple scalar field theory with fermions.}
}

The spacetime geometry of gravitating lumps is either anti-de
Sitter-Schwarzschild or de Sitter-Schwarzschild.   Dymnikova has discussed 
the global structure of de Sitter-Schwarzschild  spacetime, supposing
appropriate external matter distribution \cite{Dymnikova}.  The spacetime 
geometry of the gravitating fermionic lumps in Ref.\ \cite{Daghigh1} is  
anti-de Sitter-Schwarzschild.  The mass of such lumps is negative.  
In this paper we shall investigate fermionic lumps 
with positive mass in flat spacetime.  The equations in flat spacetime 
become simple and we expect to obtain similar solutions in curved spacetime
where the geometry is de Sitter-Schwarzschild.  The exploration of the 
de Sitter-Schwarzschild 
 spacetime is reserved for future investigation.   One may explore such 
objects in the universe at various scales and epochs.

\vskip 1cm 
In field theories with continuous global symmetries, non-dissipative 
solutions with finite energy may exist.  
These solutions are called non-topological solitons 
\cite{soliton1, soliton2, Qball}. 
The model studied in this paper is very similar to a kind of 
non-topological soliton called Fermi balls, given in Ref.\ \cite{Campbell}.
In Fermi balls, fermions have zero rest mass and their effective mass 
is due to their Yukawa coupling with a scalar field.  In such a 
model, fermions energetically have a tendency to localize at 
the boundary wall region (domain wall) where their effective 
mass becomes zero.  In our model, however, fermions have a nonzero 
rest mass and their effective mass reaches a minimum at 
the core region of the lump causing the fermions to localize 
at this region.

\ignore{
We stress that fermionic lumps described in the present
paper exist when there appears a false vacuum and there are fermions
coupled to the relevant scalar field.  Recently it has been shown that such
a false vacuum  appears in the early universe in the standard 
Einstein-Weinberg-Salam theory of electroweak and gravitational 
interactions, if the universe has a spatial section $S^3$ as in the 
closed Friedmann-Robertson-Walker universe \cite{Emoto}.  In the early
stage nontrivial gauge fields yield the false vacuum in the Higgs field,
to which quarks and leptons have the Yukawa couplings.  In other 
words gravitating fermionic lumps may be copiously produced in the
framework of the standard  model.
}

\vskip 1cm 
In section 2, we set up the problem and introduce useful
variables in terms of which the field equations are written.  
In section 3, we investigate the stability of the 
lumps with a positive mass analytically.\ignore{It illustrate 
how such lumps become possible
when the scalar field potential has false vacuum configuration.}
In section 4, the behavior of the solutions is investigated 
analytically inside the lump, numerically in the boundary wall region,
and analytically outside the lump.  More details of the solutions
are given in section 5, with a focus on the dependence of
the  solutions on various parameters of the model.   
It is seen how the solutions change as 
the energy scale of the model is lowered.  In Section 6, we 
investigate the numerical solutions in the transition region 
with the fermion density having a continuous transition to zero. 
The summary and conclusions are given in Section 7.

\sxn{Model with  Dirac fermions}

The system that we are considering is a real scalar field coupled with a
fermion field whose Lagrangian is given by 
\beeq
\mathcal{L}=i\bar\psi \gamma^{\mu} \partial_{\mu} \psi 
- m \bar\psi \psi + \frac{1}{2} \partial^\mu\phi
\partial_\mu\phi -V[\phi] - g \phi \bar\psi \psi ~.
\label{model1}
\eneq
We take 
\begin{eqnarray}
V[\phi] \, = \frac{\lambda}{4} (\phi-f_2)
\Bigg\{ \phi^3-\frac{1}{3}(f_2+4f_1)\phi^2
  - \frac{1}{3}f_2(f_2-2f_1)(\phi+f_2) \Bigg\}  ~. 
\label{potential1}
\end{eqnarray}
This potential has two non-degenerate minima at $f_1$ and \(f_2\).  
For further details on the potential refer to Ref.\ \cite{Daghigh1}.  
In this paper, we
employ the natural unit $\hbar=c=1$.

\ignore{\begin{figure}[tb]
\begin{center}
\includegraphics[height=5cm]{potential1.eps}
\end{center}
\caption{A scalar field potential $V[\phi]$ with two minima.  
$\phi \rho_S = g\phi\la \psibarb \psi\ra$  at the core is also displayed
for $g \la \psibarb \psi\ra > 0$.   In the lump solutions described 
in the present paper $g |f_1| \la \psibarb \psi\ra$ is much larger than 
the energy density of the false vacuum at $f_1$.}
\label{potential}
\end{figure}
}

As we shall see below, the scalar field $\phi$ is approximately
constant inside the lump so that fermions inside the lump may be
treated as a uniform Fermi gas.  The effective mass of the fermion 
is $m_\eff = m + g \phi_{\rm inside}$.  Depending on the density $\rho_n$ 
and effective mass $m_\eff$ of fermions,  the Fermi gas can 
be either relativistic or 
nonrelativistic.   Given $\rho_n$ and $m_\eff$, 
 $\la \psibar \psi \ra = \rho_0$ inside the lump is determined.  
Let $p_F$ be the Fermi momentum of a degenerate Fermi gas.  Then
\beqn
&&\hskip -1cm
\rho_n = {p_F^3 \over 3 \pi^2 }~,\cr
&&\hskip -1cm
\rho_0 = {m_{\eff}  \over 2 \pi^2 } 
\Big\{p_F (p_F^2+m_{\eff}^2)^{1/2} -
m_{\eff}^2 \log{p_F+ (p_F^2+m_{\eff}^2)^{1/2} \over m_{\eff} c} \Big\}~.
\label{density1}
\eeqn
Note that $\rho_0 \sim \rho_n$ if fermions are nonrelativisitic.
The energy-momentum tensors 
$T_{\mu\nu}$ for spherically symmetric, static configurations
are 
\beqn
&&\hskip -1cm 
T_{00} =\frac{1}{2}{\phi'}^2
+V[\phi]+ {\cal E}_f  ~, \cr
\noalign{\kern 8pt}
&&\hskip -1cm 
T_{11} = \frac{1}{2}{\phi'}^2
-V[\phi]+  P_f~, \cr
\noalign{\kern 8pt}
&&\hskip -1cm 
T_{22} = T_{33}= -\frac{1}{2}{\phi'}^2
        -V[\phi]+ P_f~, \cr
\noalign{\kern 8pt}
&&\hskip -1cm 
\hbox{others}  = 0 ~~, 
\label{em-tensors}
\end{eqnarray}
where  primes represent \(r\) derivatives.  Here
\beeq
{\cal E}_f = {1 \over 8\pi^2 } \Big\{p_F (2 p_F^2+m_{\eff}^2 )
(p_F^2+ {m_{\eff}^2})^{1/2} -
({m_{\eff})^4}\log \Big[ {{p_F+ (p_F^2+m_{\eff}^2)^{1/2}}
\over {m_{\eff}}} 
\Big] \Big\},
\label{relativistic energy}
\eneq
and
\beeq
P_f = {1 \over 8\pi^2 } \Big\{p_F ({2 \over 3} p_F^2-m_{\eff}^2 )
(p_F^2+ {m_{\eff}^2})^{1/2} +
({m_{\eff})^4}\log \Big[ {{p_F+ (p_F^2+m_{\eff}^2)^{1/2}}
\over {m_{\eff}}} 
\Big] \Big\}
\label{relativistic pressure}
\eneq
are the energy density and the pressure density of a Fermi gas, respectively.
\ignore{For a relativistic Fermi gas where $p_F \gg m_{\eff}c$, 
\beeq
\rho_0  \sim {m_{\eff} c p_F^2 \over 2 \pi^2 \hbar^3}=
{m_{\eff}c \over 2 \hbar} \Big( {3 \rho_n \over \pi}\Big)^{2/3}  ~.
\label{cond6}
\eneq 
In the ultrarelativistic limit ($p_F \ll m_{eff} c$) the energy-momentum tensors 
$T_{\mu\nu}$ for spherically symmetric, static configurations
are 
\beqn
&&\hskip -1cm 
T_{00} =\frac{1}{2}{\phi'}^2
+V[\phi]+ g\phi \rho_0 + {\pi^2 \hbar^3 \over c (m+g\phi)^2}\, 
\rho_0^{2}  ~, \cr
\noalign{\kern 8pt}
&&\hskip -1cm 
T_{11} = \frac{1}{2}{\phi'}^2
-V[\phi]-  g\phi \rho_0 
+ {\pi^2 \hbar^3 \over c (m+g\phi)^2}\, \rho_0^{2}  ~, \cr
\noalign{\kern 8pt}
&&\hskip -1cm 
T_{22} = T_{33}= -\frac{1}{2}{\phi'}^2
        -V[\phi]-  g\phi \rho_0 
+ {\pi^2 \hbar^3 \over c (m+g\phi)^2}\, \rho_0^{2}  ~, \cr
\noalign{\kern 8pt}
&&\hskip -1cm 
\hbox{others}  = 0 ~~, 
\label{em-tensors}
\end{eqnarray}
where  primes represent \(r\) derivatives.}
$T_{00}$ is the 
energy density of the lump which can be used to calculate 
the total mass of the lump
\begin{equation}
M=\int_0^r 4\pi r^2  dr \, T_{00} ~~.
\label{metric2}
\end{equation}  
 The equation of motion for the scalar field is 
\begin{equation}
-\frac{1}{r^2}\frac{\partial}{\partial r}
\Bigl({r^2}\phi'\Bigr)+V'[\phi]+ g\rho_S(r)=0
\label{scalar1}
\end{equation} 
where $\rho_S(r) = \la \psibar \psi \ra (r)$, and $V'[\phi]$ is the derivative of $V[\phi]$ with respect to $\phi$.

In order to solve Eq. (\ref{scalar1}) we adopt an approximation
\beeq
\rho_S(r) = \rho_0 \theta(R_1 - r) ~,
\label{f-radius}
\eneq
and we divide the space into three regions:

\bigskip
\centerline{
\begin{tabular}{l l}
I. & The inside region ($0 \le r \le R_1$ ).\\
II. & The boundary wall region ($ R_1 \le r \le R_2$ ).\\
III. & The outside region ($R_2 \le r < \infty $).\\
\end{tabular}
}

\bigskip

\begin{figure}[htb]
\begin{center}
\includegraphics[height=5cm]{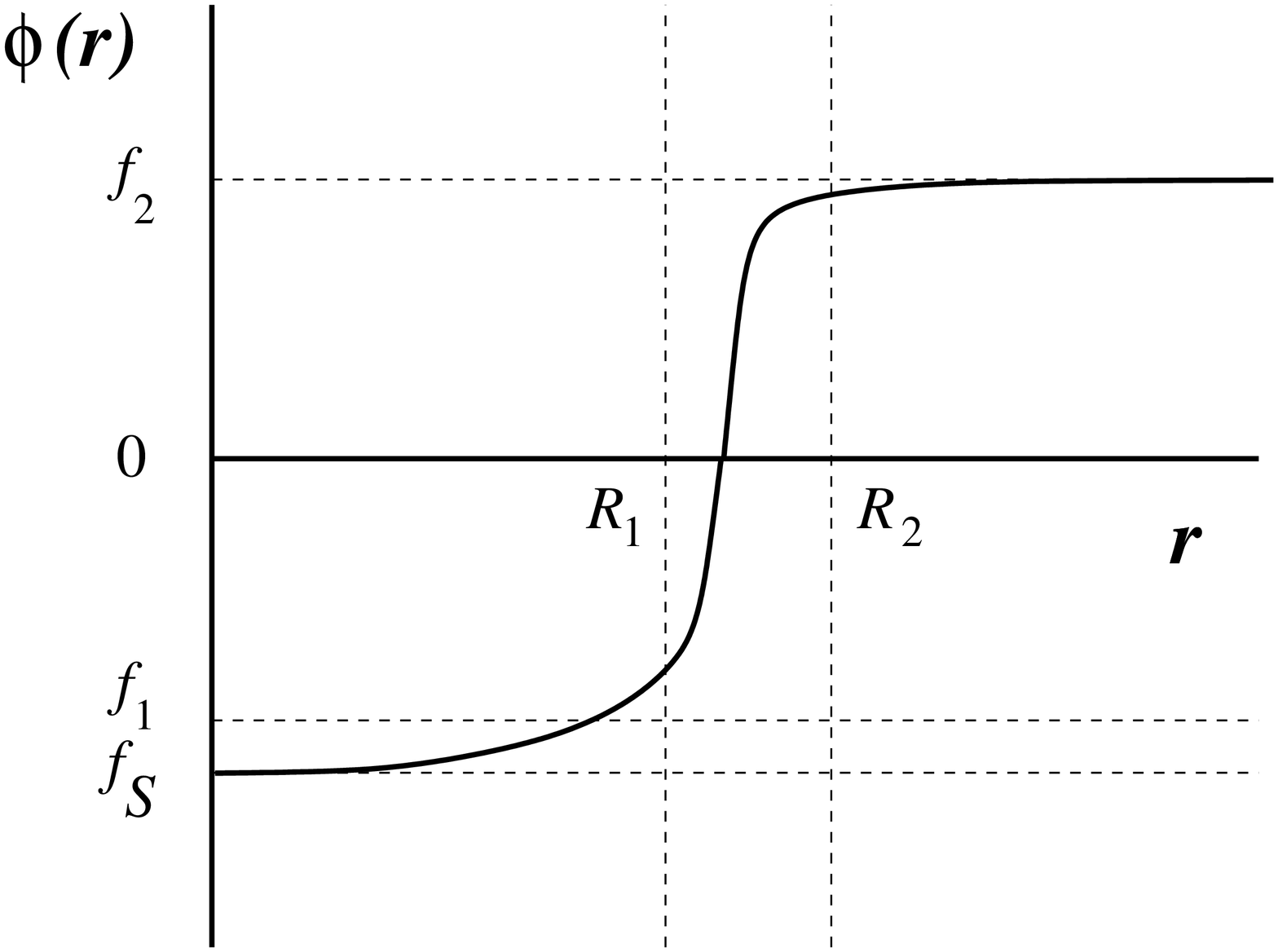}
\end{center}
\caption{Schematic behavior of $\phi(r)$.  In the lump solutions 
$|f_1 - f_S| \ll |f_1|$ and $R_2-R_1 \ll R_1$.}
\label{phi-1}
\end{figure}

\noindent
In region I inside the lump, $\rho_S =\rho_0$ 
and $\phi(0)$ is very close to, but still greater than the location $f_S$ of 
the minimum of $V[\phi] + g\phi \rho_0$: $\phi(0) = f_S + \delta \phi(0)$
with 
$0 < \delta\phi(0)/|f_S| \ll 1$.  In the inside region,
$\phi(r)$ varies little from 
$f_S$ so that the equation of motion for $\phi$ can be 
linearized.  In region II, $\rho_S=0$.  In this region the
field varies substantially and the full nonlinear equation
must be solved numerically.  In region III, $\rho_S=0$ and $\phi \sim f_2$. 
In this paper we focus on the case in which
$|f_1|\sim f_2 \sim f \equiv \onehalf (|f_1| + f_2)$, 
$f_r = (f_2 - |f_1|)/f \ll 1$  and 
$f_1-f_S \ll f$ so that the linearization of Eq. (\ref{scalar1}) is valid.
In Section 6, we will relax 
the third restriction by using a continuous 
transition from $\rho_S=\rho_0$ to $\rho_S=0$ in region II.    
Schematic behavior of  a solution $\phi(r)$ is displayed in 
Fig.\  \ref{phi-1}.

\sxn{Stability of lumps}

As we mentioned earlier, the scalar field inside the lump (region I) 
is approximately constant.  This allows us to treat the fermions
inside the lump as a uniform Fermi gas. 
In this region, the Yukawa interaction $g \phi \psibar \psi$ 
generates an additional
linear interaction $g\rho_0 \phi$ for the scalar field.  
The total potential $V[\phi] + g \rho_0 \phi$ has a
minimum at $f_S <0$.  
Let $R$ be the radius of the lump.
To see how false vacuum lumps with fermions become possible 
from energetics considerations, we suppose that
the total fermion number $N_F = 4\pi R^3 \rho_n /3$ 
is conserved and therefore fixed.
It is supposed that $V[f_1] \equiv \ep_0 > 0$ but 
the energy density at the minimum  $V[f_S] + g \rho_0 f_S =  \ep$ 
becomes negative for the lump solution.
$f_S$ and $\ep$ depend on $\rho_n$ or $R$.
Inside the lump $r < R$, the scalar field assumes $\phi \sim f_S$ whereas
outside the lump $\phi = f_2$.  
\ignore{A degenerate nonrelativistic fermion
gas has energy density ${\cal E}_f = mc^2 \rho_n + A \rho_n^{5/3}$ where
$A= 3 (3\pi^2)^{2/3} \hbar^2 / 10 m$.}
The total energy of the lump is
approximately given by 
\beeq
E(R) = \Big\{ {\cal E}_f(\rho_n, m_\eff) + \ep (\rho_0) \Big\}  \, V
 + S \sigma + E_{f, \rm wall}(R)~,
\label{energy1}
\eneq
where $\sigma$ is the surface tension resulting from varying $\phi$ in
the boundary wall region.  $V=4\pi R^3/3$ and $S=4\pi R^2$ are the 
volume and the surface area of the lump, respectively.  
$\rho_0 = \rho_0(\rho_n, m_\eff)$ as related
by (\ref{density1}).  The energy density of a 
degenerate Fermi gas ${\cal E}_f(\rho_n, m_\eff)$ is 
given in (\ref{relativistic energy}).
$E_{f, \rm wall}(R)$ is the contribution to the
energy from fermions localized in the  boundary wall region.
$E_{f, \rm wall}(R)$ has been
estimated in Ref.\ \cite{Arafune} to be about 
$4\sqrt{\pi}N_{f, \rm wall}^{3/2}/3\sqrt{S}$ for $m=0$,
where $N_{f, \rm wall}$ is the number of fermions confined in the boundary
wall region. 
In this paper, we shall focus on the case 
$N_{f, \rm wall} \ll N_F$.

\ignore{-----------------------------------------------------------------------------------------
{\bf April 23, 2004:  Yutaka I did the necessary corrections to the following Paragraphs.  I added the relativistic Fermi gas.  I did lots of changes to this part of the paper.  Please read it carefully.}}

Using Eq.\ (\ref{relativistic energy}), the energy density of a 
degenerate Fermi
gas can be found in the nonrelativistic limit where 
$p_F \ll m_{\eff}$, and in the ultrarelativistic 
limit where  $p_F \gg m_{\eff}$. In the nonrelativistic limit
\beeq
{\cal E}_f \approx m_{\eff} \rho_n + {A \over m_{\eff}} \rho_n^{5/3},
\label{nonrelativistic}
\eneq
where
$A= 3 (3\pi^2)^{2/3} / 10$, and in the ultrarelativistic limit  
\beeq
{\cal E}_f \approx B \rho_n^{4/3},
\label{ultrarelativistic}
\eneq
where $B=  (9 \pi)^{2/3}/4$.

For fixed $N_F$,    
$\ep (\rho_0) \sim \ep_0$ for $\rho_n \sim 0$ (as $R \go \infty$).
Therefore, 
$E(R) \sim (4\pi \ep_0 /3) R^3$ for large $R$ in both nonrelativistic 
and ultrarelativistic cases.
For large $\rho_n$ (as $R \go 0$), the total potential is approximated by
$(\lambda/4) \phi^4 + g \rho_n \phi$ near the minimum so that
$\ep  \sim - {3\over 4} \lambda^{-1/3} (g \rho_n)^{4/3}$.   
In the case of a lump with a nonrelativistic Fermi
gas the ${\cal E}_f$-term dominates over the $\ep$-term 
for small $R$ and
$E(R) \sim A' R^{-2}$, where $A' >0$.  On the other hand, for a 
lump with an ultrarelativistic Fermi gas  the ${\cal E}_f$-term can 
only dominate over the $\ep$-term 
for small $R$ if $B>{3\over 4} \lambda^{-1/3} g^{4/3}$ or
\beeq
{g^4 \over \lambda} > 3 \pi^2.
\label{cond0}
\eneq
If Eq.\ (\ref{cond0}) is satisfied, then $E(R) \sim B' R^{-1}$ with $B' > 0$.  
  Therefore, with Eq.\ (\ref{cond0}) 
satisfied, the total energy $E(R)$ of a false vacuum lump filled with either ultrarelativistic or nonrelativistic Fermi
gas has a minimum at a certain radius $\Rbar$, where $\Rbar$ is the size of the lump.

 \ignore{Further the above fermion
lump configuration need to be energetically favored over a configuration
in which fermions live in the scalar configuration $\phi \sim f_2$.
These conditions can be satisfied if the mass $m$ of the fermion
in the ${\cal E}_f$ differs from the mass $m_0$ in the vacuum 
$\phi = f_2$.}

\ignore{In the examples of the lump solutions below we shall have 
$g \rho_n |f_S| \gg V[f_S]$.  In other words $|\ep| \sim g \rho_n |f_S|$.} 

Now let us explore some of the conditions and restrictions 
on the lump solutions in this paper.
For the lump solutions to become possible we need to have $\ep < 0$.  
Note that $\ep < 0$ implies that
\beeq
\rho_0 > {\ep_0 \over gf} \sim {\lambda\over g} \, f^3 f_r    
~~.
\label{cond1}
\eneq 
In the examples of the lump solutions below, $f_S$ is close to $f_1$,
and it is useful to introduce the parameter $h_S = (f_1 - f_S)/f \ll 1$.
$\rho_0$ and $h_S$ are related by
\beeq
\rho_0 \sim  {2 \lambda\over g} \, f^3 h_S    
~~.
\label{cond2}
\eneq
Equations (\ref{cond1}) and (\ref{cond2}) imply that
\beeq
2 h_S > f_r    
~~.
\label{cond3}
\eneq
Also in this paper $|\ep| \gg \ep_0$.  In other words 
\beeq
|\ep| \sim gf \rho_0~.
\label{cond4}
\eneq
First we consider a lump filled with nonrelativistic Fermi gas.
As we mentioned earlier, the nonrelativistic 
approximation is justified 
if $m_{\eff}\gg p_F$, which implies that
\beeq
m_{\eff} \gg \rho_0^{1/3}  \sim
\Big( {\lambda h_S \over g^4} \Big)^{1/3} \,g f ~.
\label{cond5}
\eneq
  We would like to have a lump 
with positive energy;
${\cal E}_f + \ep > 0$.  This leads to
\beeq
m_{\eff} > { |\ep| \over \rho_n}\sim gf~~.
\label{cond6}
\eneq
If the conditions in Eqs. (\ref{cond4}) and (\ref{cond5}) are satisfied, we may 
have a false vacuum lump filled with nonrelativistic Fermi gas with positive mass.  
Now let us consider a lump filled with ultrarelativistic Fermi gas.
 The ultrarelativistic approximation is justified 
if $m_{\eff}\ll p_F$, which implies that 
\beeq
m_{\eff} \ll \rho_0^{1/3} \sim  \Big( { \lambda h_S \over g^4}\Big)^{1/3} \, gf  ~.
\label{cond7}
\eneq 
For this case also we would like to have a lump 
with positive energy;
${\cal E}_f + \ep > 0$.  This leads to
\beeq
m_{\eff} <   
{\pi \rho_0 \over |\ep|^{1/2}} \sim 
 {{2 \pi \lambda h_S }\over {g^4}}  (gf)^{5/2}~~.
\label{cond8}
\eneq
If the conditions in Eqs. (\ref{cond7}) and (\ref{cond8}) are satisfied, we may
have a false vacuum lump filled with ultrarelativistic Fermi gas which has a positive mass. 
\vskip .5cm

Note that if the fermion rest mass $m \gord gf$, then the effective mass of 
the fermion inside the lump, which is $|m-gf|$, is smaller than the 
effective mass of the fermion outside the lump, which is $m+gf$.  
This condition is satisfied for both cases of lumps with 
nonrelativistic and ultrarelativistic Fermi gas.  This is 
the reason why fermions energetically favor to reside at 
the core of the lump.  The situation will change when 
the fermion rest mass $m \ll gf$ or $m=0$.  These scenarios 
have been studied in the models suggested in
Refs.\ \cite{Daghigh1} and \cite{Campbell}.

\vskip .5cm

The stability of a lump configuration against deformation from 
spherical shape is examined in the
following way.  In Eq. (\ref{energy1}), the surface tension 
energy and the energy from fermions localized at the boundary 
region $E_{f,wall}(R)$ depend only on the surface area of the 
lump and not on its volume.  On the other hand, it is clear from 
Eqs. (\ref{energy1}),  (\ref{nonrelativistic}), (\ref{ultrarelativistic}), 
and (\ref{cond4}) 
that the volume energy will increase if we decrease the volume 
of the lump by deforming it from spherical shape for both 
nonrelativistic and ultrarelativistic cases.

\ignore{The stability of a lump configuration 
against fission is examined in the
following way.  First, we suppose that a small amount of fermions, say
$N_1$ fermions ($N_1 \ll N_F$), are emitted to the space infinity from
the lump.  The density of fermions inside the lump is unchanged whereas
the size $R$ becomes smaller.}

\ignore{Next, we consider the process in which the initial lump breaks up
into two lumps with the fermion numbers $N_1$ and $N_2$ ($N_1+N_2= N_F$).
In this case ...}    


\sxn{Behavior of the solutions}

The procedure to solve Eq. (\ref{scalar1}) using 
the assumption described in the previous section is as follows:

\noindent
{\bf Region I}

First we determine $f_S$ using
\beeq
V'[f_S] + \rho_0 = 0 ~~.
\label{fS}
\eneq
$f_S$ and $f_1$ are close to each other compared to the length 
scale $f$.   Denoting
$\phi(0)$ by $\phi_0$,  we have $0 < \phi_0 - f_S \ll |f_S|$.
\ignore{\beqn
\phi(r) &=&\phi_0+\phi_2 r^2+\cdots ~~, 
 \hskip 1cm \phi_2= \frac{1}{6}
  \bigl( V'[\phi_0]+ \rho_0 \bigr) ~.
\label{origin1}
\eeqn}

The equation for $\phi(r)$ can be linearized with
$\delta \phi(r) = \phi(r) - f_S $.  In terms of $z \equiv r\omega$,
\beqn
\Bigg\{ {d^2\over dz^2} 
+ \frac{2}{z} {d\over dz}
- 1\Bigg\} \, \delta\phi = 0 ~,
\label{scalar4}
\eeqn
where $\omega^2 = V''[f_S]$.  The solution to Eq. (\ref{scalar4}) 
regular at $r=0$ is
\beqn
\delta \phi(r) = \delta \phi (0)
\frac{\sinh z}{z}.
\label{phi1}
\eeqn
As one can see, the ratio $\delta\phi(r)/\delta \phi(0)$ grows exponentially   
as $e^z/2z$.  At the end of region I, $\delta\phi/|f_S|$ needs
to be very small for the linearization to be valid.  
\ignore{We shall soon see that a solution with lump structure appears for
$z \gg 1$ {\bf (??)} with a particular choice of $\delta \phi(0)$.}
The deviation from $f_S$ at the origin, $\delta\phi(0)$, needs to be
very small to have  an acceptable solution.  

The ratio of $\delta \phi'(r)$ to $\delta \phi(r)$ is given by
\beeq
{\delta \phi'(r) \over \delta \phi(r)}
= \omega\left(\frac{\cosh z}{\sinh z}-\frac{1}{z}\right)\sim \omega~,
\label{phi2}
\eneq
for $z \gg 1$.
\ignore{\beeq
{\delta \phi'(r) \over \delta \phi(r)}
\sim \omega\left(1-\frac{1}{z}\right)\sim \omega~.
\label{phi3}
\eneq
}

\begin{figure}[tb]
\begin{center}
\includegraphics[height=5cm]{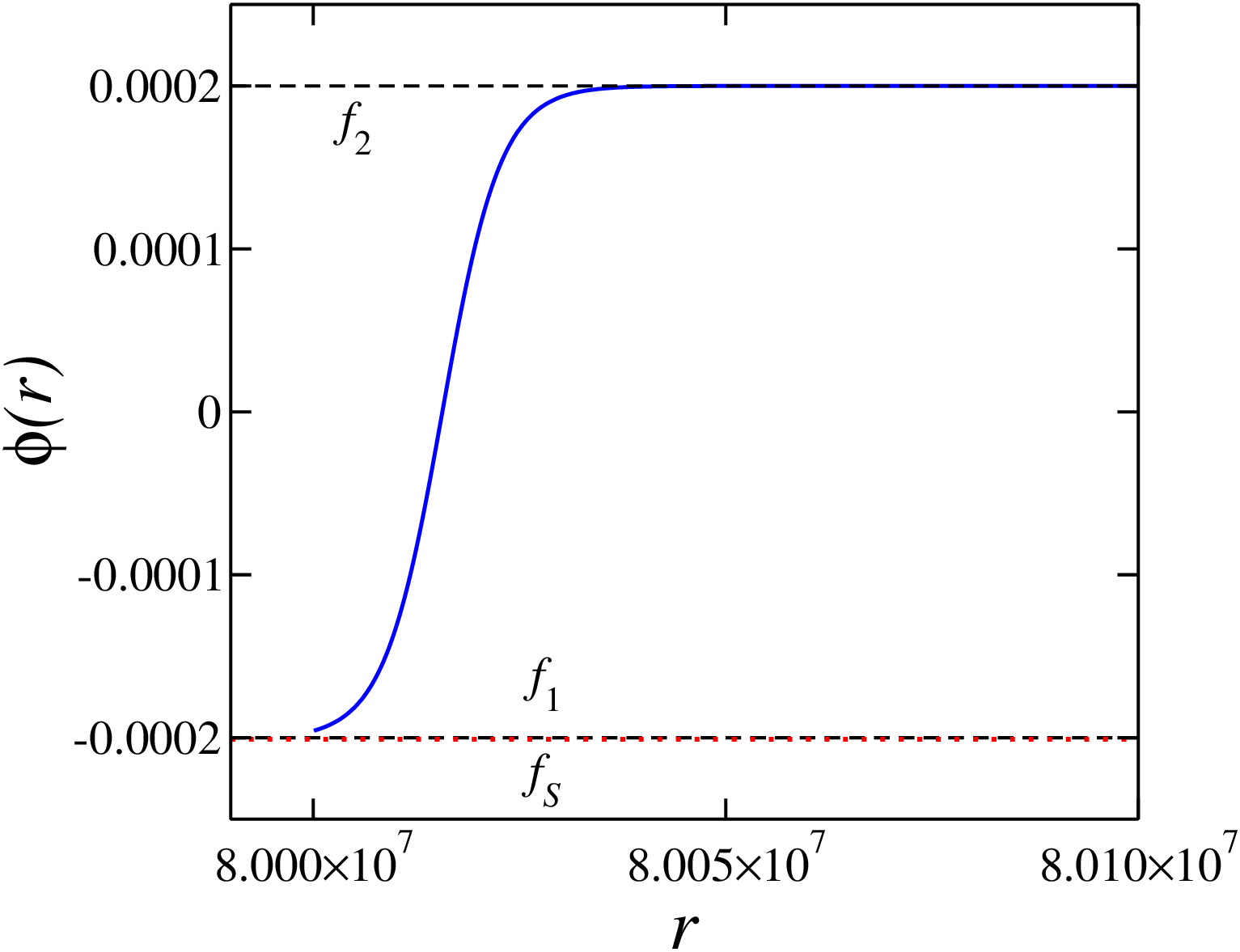}
\end{center}
\caption{$\phi(r)$ of a solution with $f/M_\Pl=0.0002$, $f_r=0.0002$, 
$\lambda=g=1$, $h_S=(f_1-f_S)/f=0.005$ and $R_1/l_\Pl=8\times 10^7$.  
$\phi$ and $r$ are in the units of $M_\Pl$ and $l_\Pl$, respectively.}
\label{phi-r-flat}
\end{figure}

\begin{figure}[tb]
\begin{center}
\includegraphics[height=5cm]{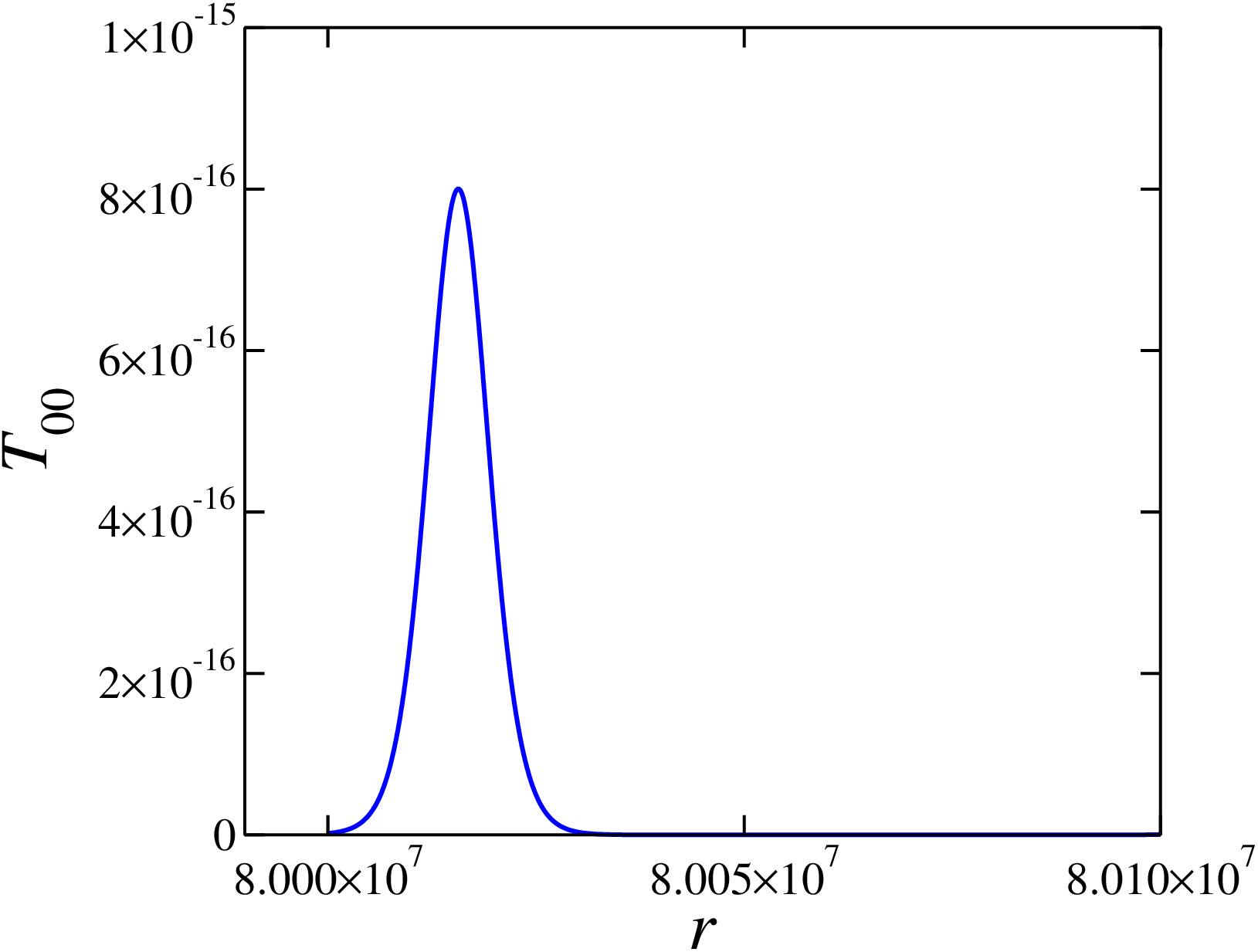}
\end{center}
\caption{The energy density $T_{00}$  for a solution with $f/M_{\Pl}=
0.0002 , f_r= 0.0002 , \lambda = g = 1$, $h_S=(f_1-f_S)/f=0.005$ and
$R_1/l_{\Pl}=8 \times 10^7$. 
$T_{00}$ and $r$ are in the units of $M_{\Pl}^4$ and $l_{\Pl}$, 
respectively.  For $r<R_1$, $T_{00} \sim \ep  +m \rho_n
+ \myfrac{3 (3\pi^2)^{2/3}}{10 m} \, \rho_n^{5/3} >0$.}
\label{T_00-r-flat}
\end{figure}

\bigskip
\noindent
{\bf Region II}

In region II, the equation must be solved numerically because $\phi(r)$ 
varies significantly.
Nontrivial lump solutions become possible by fine 
tuning the value of $\delta\phi(R_1)$.

How to find the numerical solutions is summarized as follows.
First we choose $R_1$, $\delta \phi(R_1)$, and
$f_S$.  $\delta\phi'(R_1)$ is evaluated from the analytic 
solution in region I.  
Using the boundary conditions
$\delta\phi(R_1)$ and $\delta \phi'(R_1)$, we solve the equation numerically 
in this region. The width of the boundary wall region, $w = R_2 - R_1$, 
is approximately given by $1/\sqrt{\lambda} \, f$ \cite{Hosotani1}.

One example of the lump solutions in region II is displayed 
in Fig. \ref{phi-r-flat}.  In this example, we choose the 
input parameters $g=1$,
$\lambda=1$, $f/M_\Pl = 0.0002$, $f_r=\Delta f/f = 0.0002$ and
$h_S=(f_1- f_S)/f =0.005$.  The resulting output parameters
are:
$\rho_0 l_\Pl^3  =8.0593\times 10^{-14}$,
 $\epsilon  /M_\Pl^4  =-1.5944\times10^{-17}$, $|\ep|/\ep_0 = 74.738$ and
$\omega / M_\Pl =2.8495\times10^{-4}$.   We note that $\rho_0/f^3 = 0.0101$ and
$\ep/f^4= - 0.00997$. For $R_1 /l_{\Pl} =8\times 10^7$ ($R_1\omega =
2.2796\times10^4$), we find a solution with 
$\delta\phi(R_1)/M_\Pl=5.1491304 \cdots \times 10^{-6}$.  
In this example, the value of
$\delta\phi$ at the origin ($r=0$) is found from Eq. (\ref{phi1}) to be
$1.562 \times 10^{-9901}$, which explains why one cannot numerically
integrate $\phi(r)$ starting from $r=0$ to $R_1$.
A small discontinuity
in $\phi''$ appears at $r=R_1$ due to the discontinuous change in
$\rho_S(r)$.  


\vskip .5cm
Suppose that the fermion mass is $10^{-3}M_\Pl$.
Then for the given example above, we have a false vacuum 
lump filled with nonrelativistic Fermi gas which has a total mass of
\beeq
M \approx \Big\{ \ep (\rho_n) +m_{\eff}\rho_n+  
\myfrac{3 (3\pi^2)^{2/3}}{10 m_{\eff}} 
\, \rho_n^{5/3} \Big\}  \, {4\pi
R^3\over 3} =1.4 \times 10^{8} M_\Pl.
\label{energy2}
\eneq
For the same given example, let us take the fermion mass to 
be $10^{-19}M_\Pl$, which is approximately equal to proton mass.
In this case, we would have a false vacuum lump filled 
with ultrarelativistic Fermi gas with a total mass of
\beeq
M \approx \Big\{ \ep (\rho_0) +  \myfrac{(9\pi)^{2/3}}{4} 
\, \rho_n^{4/3} \Big\}  \, {4\pi
R^3\over 3} =1.4 \times 10^{37} M_\Pl.
\label{energy3}
\eneq
In both of these examples, the energy contribution from surface tension 
$\sigma$ is small ($\sim 10^6 M_\Pl$)
and can be ignored.

The behavior of energy density $T_{00}$ is displayed in 
Fig. \ref{T_00-r-flat}.  The energy density has one sharp peak 
associated with the rapid variation of $\phi$.

The field $\phi(r)$ approaches $f_2$ for $r > R_2$.  In the numerical
integration, $R_1$ and $f_S$ are kept fixed while  $\delta\phi(R_1)$ is
varied.   If $\delta\phi(R_1)$ is chosen to be slightly smaller, $\phi$
comes back toward, but cannot reach $f_1$, eventually oscillating around
$\phi=0$ as $r$ increases.  
If $\delta\phi(R_1)$ is taken to be slightly
bigger, then $\phi(r)$ overshoots $f_2$, heading for $+\infty$ as $r$
increases.  With just the right value, $\phi(r)$ approaches to $f_2$ as 
$r\rightarrow \infty$ in region III.  Shell solutions 
can appear in which $\phi$ goes back to $f_1$ at $r=\infty$.

\bigskip
\noindent
{\bf Region III}

In region III, $\phi(r)$ is very close to  $f_2$.  
The equation for $\phi(r)$ can be linearized with
$\delta \phi(r) =f_2 -  \phi(r)$.  The linearized equation is 
the same as Eq. (\ref{scalar4}) with $z \equiv r \alpha$, 
where $\alpha^2 = V''[f_2]$.  The solution in which
$\delta \phi(r\rightarrow \infty)=0$ is
\beqn
\delta \phi(r) = {{\delta \phi (R_2) \cdot {{e^{-z+z_2}}\over{z/z_2}}}}.
\label{phi4}
\eeqn
The value of $\delta \phi (R_2)$ is determined numerically.

\begin{figure}[tb]
\begin{center}
\includegraphics[height=5cm]{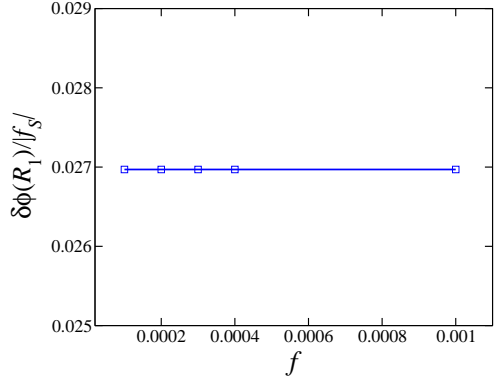}
\end{center}
\caption{The $f$ dependence of $\delta \phi (R_1)$.  $\lambda=g=1$, 
$f_r=0.0002$, 
$R_1 \omega=2 \times 10^4$ and
$h_S=(f_1-f_S)/f=0.005$ are fixed.  $f$ is in the units of $M_\Pl$.}
\label{f-flat}
\end{figure}

\sxn{Numerical Analysis in the Nonlinear Regime}

In the previous section, we solved the field equation in three different 
regions of space.
In this section, we analyze our numerical results in more detail.
\ignore{As discussed in  the previous section the boundary
between regions I and II, located at the matching radius $R_1$, is rather
arbitrary, subject only to the condition that the linearization is
accurate up to that radius.  Precise tuning is necessary for $R_1$ and 
$\delta\phi(R_1)$ to obtain a lump solution.  Technically it is easier to
keep $R_1$ fixed and adjust
$\delta\phi$ at $R_1$.   If $\delta\phi(R_1)$ is chosen too small $\phi$
comes back toward, but cannot reach $f_1$,   eventually oscillating around
$\phi=0$ as $r$  increases.  If $\delta\phi(R_1)$ is chosen too large $\phi$
overshoots $f_2$ and continues to increase.   With just the right
value of $\delta\phi(R_1)$, $\phi$ will approach to $f_2$.  
There can appear shell solutions in which $\phi$ goes back to $f_1$ at 
$r=\infty$.}  

\begin{figure}[tb]
\begin{center}
\includegraphics[height=5cm]{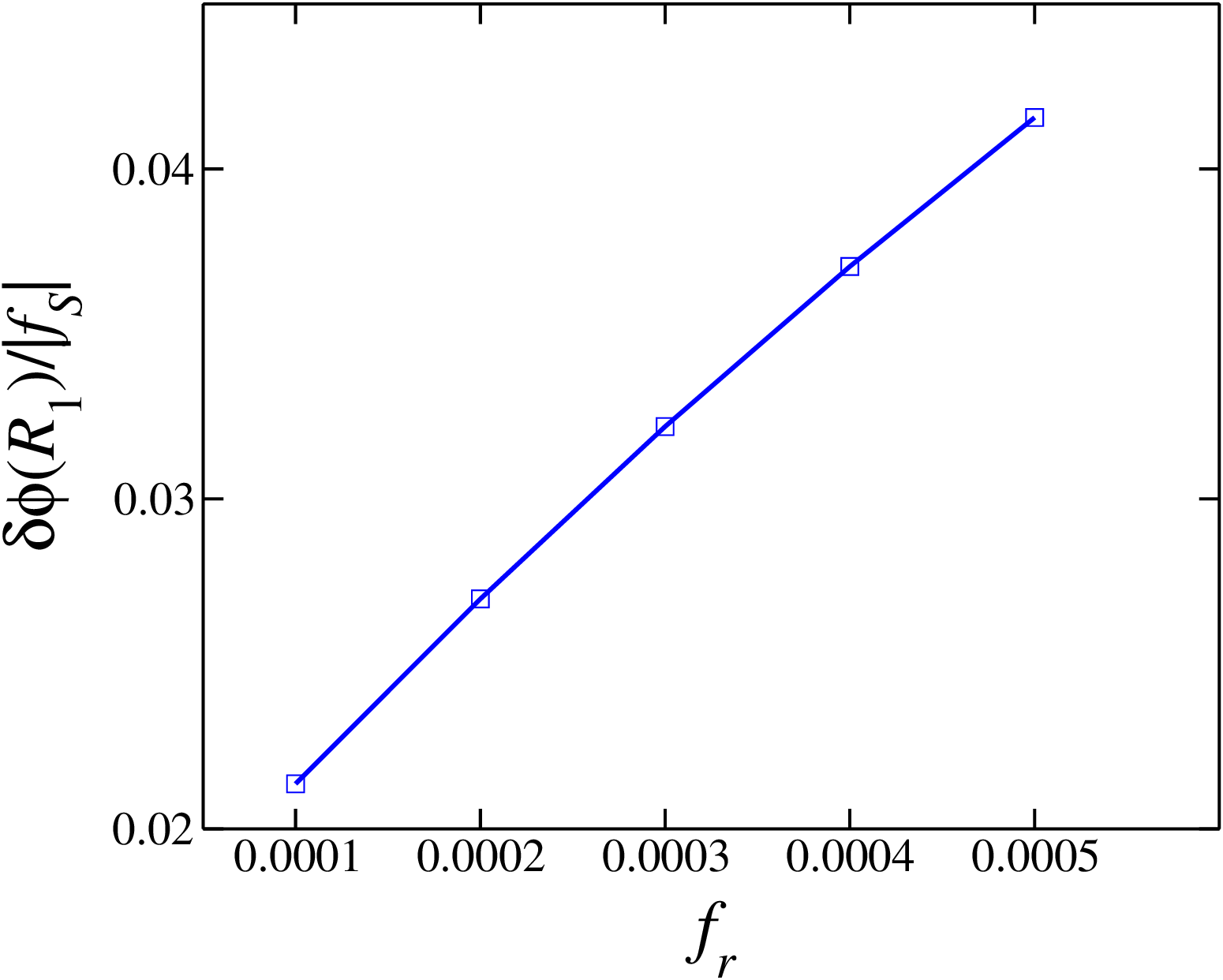}
\end{center}
\caption{The $f_r$ dependence of $\delta \phi (R_1)$.  
$\lambda=g=1$, $f/M_\Pl=0.0002$, 
$R_1 \omega=2 \times 10^4$ and
$h_S=(f_1-f_S)/f=0.005$ are fixed. }
\label{f_r-flat}
\end{figure}

\begin{figure}[tb]
\begin{center}
\includegraphics[height=5cm]{R_1-flat.eps}
\end{center}
\caption{The $R_1$ dependence of $\delta \phi (R_1)$.  
$\lambda=g=1$, $f/M_\Pl=0.0002$, $f_r=0.0002$
and $h_S=(f_1-f_S)/f=0.005$ are fixed.}
\label{R_1-flat}
\end{figure}

\begin{figure}[tb]
\begin{center}
\includegraphics[height=5cm]{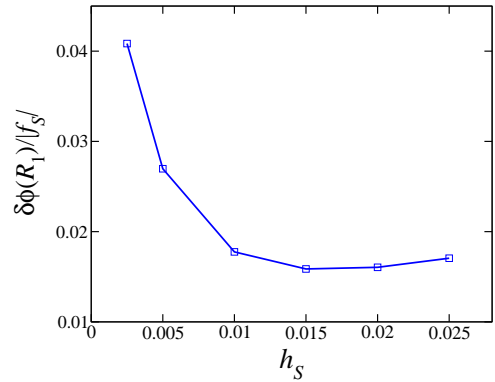}
\end{center}
\caption{The $h_S$ dependence of $\delta \phi (R_1)$.  
$\lambda=g=1$, $f/M_\Pl=0.0002$, $f_r=0.0002$
and $R_1 \omega =2 \times 10^4$ are fixed.}
\label{f_S-flat}
\end{figure}

\begin{figure}[tb]
\begin{center}
\includegraphics[height=5cm]{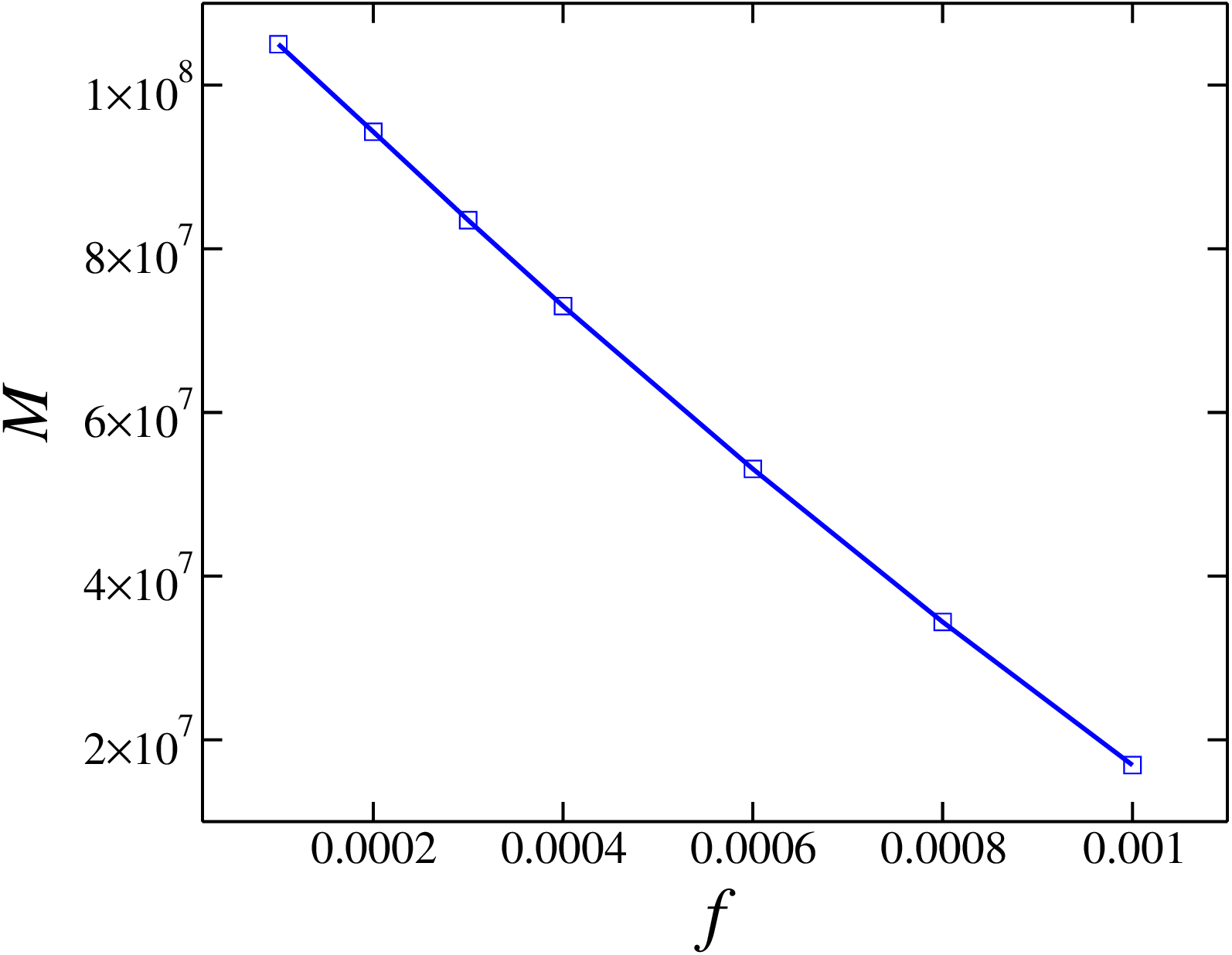}
\end{center}
\caption{$M$ versus $f$.  $f_r=0.0002$, $h_S=(f_1-f_S)/f=0.005$, 
$R_1 \omega =2 \times 10^4$ and $m_{\eff}/M_\Pl=10^{-3}$ are fixed.  
$f$ and $M$ are in the units of 
$M_{\Pl}$. 
}
\label{mass1}
\end{figure}

\begin{figure}[tb]
\begin{center}
\includegraphics[height=5cm]{M-m-flat.eps}
\end{center}
\caption{
$M$ versus $m_{\eff}$.  $f/M_{\Pl}=0.0002$, $f_r=0.0002$, 
$h_S=0.005$ and $R_1 \omega =2 \times 10^4$ are fixed.  
$M$ and $m_{\eff}$ are in the units of $M_{\Pl}$.}
\label{mass2}
\end{figure}

\begin{figure}[tb]
\begin{center}
\includegraphics[height=5cm]{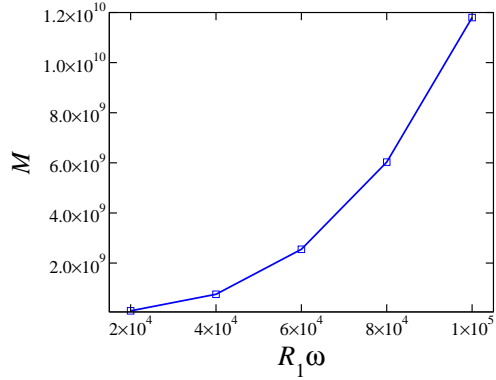}
\end{center}
\caption{
$M$ versus $R_1 \omega$.  $f/M_{\Pl}=0.0002$, $f_r=0.0002$, 
$h_S=0.005$ and $m_{\eff}/M_\Pl=10^{-3}$ are fixed.  $M$ is in the 
units of $M_{\Pl}$.}
\label{mass3}
\end{figure}

The lump solutions depend on several dimensionless parameters 
of our model such as 
$f/M_{\Pl}$,
$f_r$, $R_1\omega$, and $\rho_0 f^3$.  The dependence on $g$ and 
$\lambda$ can be absorbed by rescaling.
When the values of these
parameters are varied, the size of the resulting lump solutions also varies.
In the numerical evaluation, we have taken  
$f$ and $f_r$ in the range  between $10^{-4}$ and $10^{-3}$.  
In addition to these parameters, the fermion mass $m_\eff = m - g|f_S|$ 
inside the lump affects the total mass (energy) of the lump.

In the numerical investigation  $g$, $\lambda$, $f$,
$f_r$, $R_1 \omega$ and $h_S$ are given to find a  desired value
of $\delta\phi(R_1)$ for a solution.  
In Fig. \ref{f-flat}, $\delta \phi (R_1)/|f_S|$
is plotted as a function of $f$.  From the figure it is clear 
that the value of the ratio $\delta \phi/|f_S|$
at $R_1$ is independent of $f$.
In Fig. \ref{f_r-flat}, $\delta \phi (R_1)/|f_S|$ is plotted as a function of
$f_r$.  $\delta \phi (R_1)/|f_S|$ increases as $f_r$ increases.  The 
numerical evaluation becomes unreliable 
when $\delta \phi (R_1)/|f_S|$ becomes large and the
linearization in Eq.\ (\ref{scalar1}) becomes invalid.  
In Fig. \ref{R_1-flat}, $\delta\phi (R_1)/|f_S|$ is plotted 
versus $R_1\omega$.  As $R_1\omega$ increases,
$\delta \phi (R_1)/|f_S|$ decreases and approaches a constant ($\sim 0.016$).  
In Fig. \ref{f_S-flat}, $\delta \phi (R_1)/|f_S|$ is plotted versus $h_S$.  
In this figure, $\delta \phi (R_1)/|f_S|$ decreases as $h_S$ increases 
and then begins to increase when
$h_S$ becomes greater than approximately $0.015$.  The
reason for the increase in the value of $\delta \phi (R_1)/|f_S|$ is that at
some point $f_1-f_S$ becomes greater than $\delta\phi(R_1)$, which means
that $\phi(R_1)<f_1$.  In this situation, the solution will diverge to
negative infinity if we do not take a large enough value for
$\delta\phi(R_1)$.  

In Figs. \ref{mass1}, \ref{mass2}, and \ref{mass3}, we plot the 
mass of the lump $M$ as a function of $f$, $m_{\eff}$, and 
$R_1\omega$, respectively.  For the data represented in these 
figures, the nonrelativistic approach is justified, 
i.e. $p_F \ll m_{\eff}$.  Combining Eqs. (\ref{cond2}), 
(\ref{cond4}), (\ref{energy2}), and the fact that 
\beeq
\omega^2\sim2\lambda f^2~,
\label{omega}
\eneq 
we may find a rough estimate of the mass of the lump 
\beeq
M \sim  (-gf +  m_{\eff}) \left(
\frac{h_S}{g\sqrt{2\lambda}}\right){4\pi (\omega R)^3\over 3}~.
\label{mass-nonrel}
\eneq 
This rough estimate explains the behavior of $M$ in 
Figs. \ref{mass1}, \ref{mass2}, and \ref{mass3}.  
Similarly, using Eqs. (\ref{cond2}), (\ref{cond4}), (\ref{energy3}), 
(\ref{omega}), and the fact that 
$\rho_n\sim \pi(2\rho_0/m_{\eff})^{2/3}/3$ for an 
ultrarelativistic Fermi gas, it is possible to find a 
rough estimate of the mass of a lump filled with ultrarelativistic 
Fermi gas
\beeq
M \sim  \left( -gf +  \frac{2\pi^2 \lambda}
{g m_{\eff}} h_S f^3\right) \left(\frac{h_S}
{g\sqrt{2\lambda}}\right){4\pi (\omega R)^3\over 3}~.
\label{mass-rel}
\eneq 
Using this equation, it is easy to extract the behavior 
of the lump mass $M$ as a function of different parameters of interest.

\sxn{Continuous transition in the boundary wall region}

\begin{figure}[tb]
\begin{center}
\includegraphics[height=5cm]{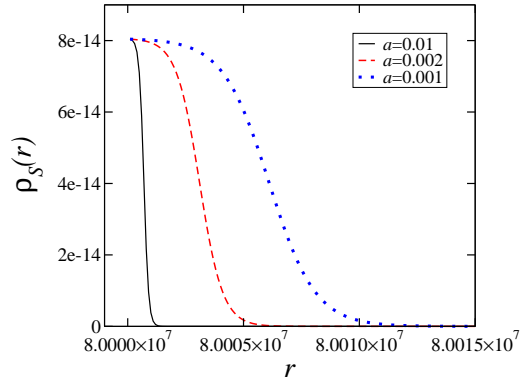}
\end{center}
\caption{$\rho_S(r)$ for $a/M_\Pl=0.01$, 
$a / M_\Pl=0.005$ and $a/M_\Pl=0.001$.  
$R_1/l_\Pl = 8 \times 10^7$ and 
$\rho_0/M_\Pl^3=8.0593\times 10^{-14}$ are fixed.  $\rho_S$ is in 
the units of $M_\Pl^3$.}
\label{rho-r-flat}
\end{figure}

\begin{figure}[tb]
\begin{center}
\includegraphics[height=5cm]{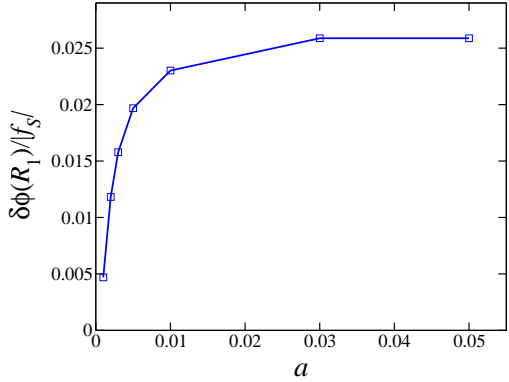}
\end{center}
\caption{$\delta \phi (R_1)/|f_S|$ versus $a$.  
$f/M_\Pl=0.0002$, $f_r=0.0002$, $h_S=(f_1-f_S)/f=0.005$ 
and $R_1\omega=2\times 10^4$ are fixed.  $a$ is in the units of $M_\Pl$.}
\label{phi-a-flat}
\end{figure}

So far we have studied the lump solutions using Eq. (\ref{f-radius}) to 
express $\rho_S(r)$.  In this approximation, a small discontinuity 
appears in $\phi''$ at $r=R_1$ due to the discontinuous change 
in $\rho_S$ at this radius.  In this section, we 
investigate the case in which $\rho_S$ makes a continuous transition from 
$\rho_0$ to $0$ in the boundary wall region (region II).  
To assure a smooth transition in this region, we 
express $\rho_S(r)$ with an equation of the form
\beeq
\rho_S(r) = \rho_0 \left\{ 1-{{1}\over {1+\exp{[a(R_1-r)+6]}}} \right\} ~~.
\label{rho-smooth}
\eneq
This equation becomes identical to Eq. (\ref{f-radius}) in 
the limit where $a\rightarrow \infty$.  In Fig. \ref{rho-r-flat}, 
we plot $\rho_S(r)$ 
for three different values of $a$.  As $a$ gets smaller, 
$\rho_S$ make a slower transition from $\rho_0$ to zero.
In Fig. \ref{phi-a-flat}, we show how $\delta \phi (R_1)$ decreases as we 
decrease the value of $a$.  In the examples shown 
in Fig. \ref{phi-a-flat}, the 
transition region from $\rho_S=\rho_0$ to $\rho_S=0$ is smaller 
than the width $w=R_2-R_1$.  
As $a \rightarrow \infty$, the value of $\delta \phi (R_1)/|f_S|$ 
in Fig. \ref{phi-a-flat} 
converges to the 
value of $\delta \phi (R_1)/|f_S|$, which we find using the approximation in
Eq. (\ref{f-radius}).  

\sxn{Summary and conclusions}
In this paper, we have shown that false vacuum 
lumps with fermions in flat spacetime may exist.  
This configuration is an interesting configuration which may 
appear in different scales and masses.  The size of these 
lumps depends on the typical energy scale of the scalar 
field potential.  
If the energy scale is very high, the size
of the lump is small.  If the energy scale is low, 
the lump may extend at a cosmic scale.  On the other hand, the mass of these 
lumps depends both on the energy scale of the scalar 
potential and the mass density of fermions inside the lump.
For lumps with large masses and sizes, the gravitational 
effects need to be included.  The spacetime structure for such 
lumps will be de Sitter- Schwarzschild.  We leave the 
investigation on de Sitter-Schwrzschild gravitating false vacuum 
lumps with fermions to a future project.  The stability of the 
solutions requires further examination as well.   

False vacuum lumps with fermions are a type of non-topological 
soliton that may be produced in the early universe due to the 
spontaneous breaking of an approximate discrete 
symmetry that occurs in a large family of field theories 
in four spacetime dimensions.  It has also been shown in Ref.\ \cite{Emoto} that 
the false vacuum lumps of the Higgs field may emerge in the early universe 
in the standard 
Einstein-Weinberg-Salam theory of electroweak 
and gravitational interactions.  In this model, the fermions are 
the quarks and leptons that are coupled with the Higgs 
field via Yukawa interaction.  It would be interesting to study 
the cosmological effects and applications of such lumps. 

\ignore{Similar to Fermi-balls \cite{Campbell}, false vacuum lumps with fermions
may be a cold dark matter candidate if they are composed of 
fermions that possess no standard model gauge charges.  
The false vacuum state may appear in the standard model of 
electroweak and gravitational interactions.  In Ref. \cite{Emoto}, it has
been shown that the false vacuum of the Higgs field emerges in the early
universe as a result of the  winding of the gauge fields in the 
$S^3 \times R^1$ Robertson-Walker spacetime.  As quarks and leptons have
relevant Yukawa couplings to the Higgs field, gravitating  lumps
with quarks and leptons at the core  may be copiously produced.
As the universe expands, the barrier separating the false vacuum from the
true vacuum disappears so that the gravitating fermionic lumps would 
become unstable, the fermions inside the lumps dissipating to the
infinity.  It is of great interest to find consequences of this
process.}

\ignore{Furthermore, in the higher-dimensional gauge theory defined 
on orbifolds,false vacua naturally appear in the gauge field 
configurations.\cite{HHHK} 
It is   curious   to investigate if the gauge interactions of fermions
produce gravitating lumps just as we have found for the Yukawa
interactions.}

\vskip .5cm

\leftline{\bf Acknowledgments}

This work was supported in part by the Natural Sciences and Engineering 
Research Council of Canada.  I would like to thank Yutaka Hosotani for 
sharing his intuition and for providing invaluable assistance in
the preparation of this paper.

\ignore{This work was supported in part by   by Scientific Grants 
from the Ministry of Education and Science, Grant Number 13135215 
 and Grant Number 13640284.  The very early stage of the investigation 
 on the subject in the paper was carried out with Masafumi Hashimoto
whose contribution is deeply acknowledged.}


\def\jnl#1#2#3#4{{#1}{\bf #2} (#4) #3}

\def\Zphys{{\em Z.\ Phys.} }
\def\jssc{{\em J.\ Solid State Chem.\ }}
\def\jpsJ{{\em J.\ Phys.\ Soc.\ Japan }}
\def\ptps{{\em Prog.\ Theoret.\ Phys.\ Suppl.\ }}
\def\PTP{{\em Prog.\ Theoret.\ Phys.\  }}

\def\JMP{{\em J. Math.\ Phys.} }
\def\NPB{{\em Nucl.\ Phys.} B}
\def\NP{{\em Nucl.\ Phys.} }
\def\PLB{{\em Phys.\ Lett.} B}
\def\PL{{\em Phys.\ Lett.} }
\def\PRL{\em Phys.\ Rev.\ Lett. }
\def\PRB{{\em Phys.\ Rev.} B}
\def\PRD{{\em Phys.\ Rev.} D}
\def\PR{{\em Phys.\ Rev.} }
\def\PRe{{\em Phys.\ Rep.} }
\def\AP{{\em Ann.\ Phys.\ (N.Y.)} }
\def\RMP{{\
em Rev.\ Mod.\ Phys.} }
\def\ZPC{{\em Z.\ Phys.} C}
\def\SCI{\em Science}
\def\CMP{\em Comm.\ Math.\ Phys. }
\def\MPLA{{\em Mod.\ Phys.\ Lett.} A}
\def\IJMPB{{\em Int.\ J.\ Mod.\ Phys.} B}
\def\cmp{{\em Com.\ Math.\ Phys.}}
\def\JPA{{\em J.\  Phys.} A}
\def\CQG{\em Class.\ Quant.\ Grav. }
\def\ATMP{{\em Adv.\ Theoret.\ Math.\ Phys.} }
\def\ibid{{\em ibid.} }
\vskip 1cm

\leftline{\bf References}

\renewenvironment{thebibliography}[1]
        {\begin{list}{[$\,$\arabic{enumi}$\,$]}  
        {\usecounter{enumi}\setlength{\parsep}{0pt}
         \setlength{\itemsep}{0pt}  \renewcommand{\baselinestretch}{1.2}
         \settowidth
        {\labelwidth}{#1 ~ ~}\sloppy}}{\end{list}}


\end{document}